# Self-Assembled Fatty Acid Crystalline Coatings Display Non-Toxic Superhydrophobic Antimicrobial Properties

*Elena Prudnikov [a], Iryna Polishchuk [a], Andy Sand [b], Hanan Abu Hamad [b], Naama Massad-Ivanir [b], Ester Segal [b*] and Boaz Pokroy [a*]*

[a*] Department of Materials Science and Engineering, Technion − Israel Institute of Technology, 32000 Haifa, Israel. E-mail: bpokroy@tx.technion.ac.il; Tel: +972-4-829-4584

[b*] Faculty of Biotechnology and Food Engineering, Technion − Israel Institute of Technology, 32000 Haifa, Israel

## Abstract

Superhydrophobcity is a well-known wetting phenomenon found in numerous plants and insects. It is achieved by the combination of the surface's chemical properties and its surface roughness. Inspired by nature, numerous synthetic superhydrophobic surfaces have been developed for various applications. Designated surface coating is one of the fabrication routes to achieve the superhydrophobicity. Yet, many of these coatings, such as fluorine-based formulations, may pose severe health and environmental risks, limiting the applicability. Herein, we present a new family of superhydrophobic coatings comprised of natural saturated fatty acids, which are not only a part of our daily diet, but can be produced from renewable feedstock, providing a safe and sustainable alternative to existing state-of-the-art. These crystalline coatings are readily fabricated via single-step deposition routes, thermal deposition or spray-coating. The fatty acids self-assemble into highly hierarchical crystalline structures exhibiting a water contact angle of ~165° and contact angle hysteresis lower than 6°, while their properties and morphology depend on the specific fatty acid used as well as on the deposition technique. Moreover, the fatty acid coatings demonstrate excellent thermal stability. Importantly these new family of coatings display excellent anti-biofouling and antimicrobial properties against *Escherichia coli* and *Listeria innocua*, used as relevant model Gram-negative and Gram-positive bacteria, respectively. We believe that these coatings have a great application potential in the fields, where other alternatives are prohibited due to safety limitations, while at the same time their usage in other regulation-free applications is not limited.

**Introduction**

Nature is replete with materials that demonstrate unique functional properties such as optical and magnetic properties for sensing, mechanical properties for improved strength, superhydrophobicity for self-cleaning and more.[1–4] These properties are achieved due to exceptional balance of the structure-function relationship.[5,6]

The phenomenon of superhydrophobicity is well known for many years now and is commonly observed in a wide range of organisms like plants and insects.[6,7,8] Superhydrophobicity often provides organisms with additional functional properties such as self-cleaning, enhanced flight capability, thermal isolation and sensory capabilities.[6,7] Superhydrophobic surfaces demonstrate water contact angles (CA) higher than 150°, and contact angle hysteresis (CAH) lower than 10° in the case that the surface exhibits self-cleaning properties as well. The surface wetting can occur under 3 states: Young state, which describes a droplet on a flat surface[9]; Wenzel state, which describes a wetting with homogeneous solid-liquid interface[10]; Cassie-Baxter state, which describes a condition when the droplet is only partially supported by the structured surface[11]. Cassie-Baxter state, which assures superhydrophobicity is achieved by the combination of intrinsic hydrophobicity of the surface and its appropriate roughness. Moreover, hierarchical structure of the surface greatly contributes to lowering the surface-liquid contact area, thereby increasing CA values.[9,11,12] Superhydrophobic surfaces are also interesting as bio-inspired artificial analogues which exhibit other functional surface properties such as water repellence, anti-fog, anti-icing, reduced adhesion and antibiofouling properties.[6,13–18]

Multiple approaches have been employed to design superhydrophobic surfaces demonstrating a necessary combination of the appropriate roughness and chemical properties such as lithography and templating methods, plasma treatments, various deposition methods, as well as completely different strategy based on slippery liquid-infused porous surfaces (SLIPS).[19,20–23] Most of these fabrication routes are complicated, expensive or limited to specific materials and final usage purposes.

Previously, we have demonstrated a bio-inspired approach to form self-cleaning superhydrophobic surfaces composed of paraffin wax crystals, which self-assemble into highly-oriented hierarchical structures thereby forming a superhydrophobic surface coating. These coatings were shown to be applied onto various types of surfaces via thermal deposition.[24,25] Time-dependent tuning of the coating's wetting properties and dependence of the superhydrophobic performance on molecular weight of the various wax crystals or their combination was studied.[24,26] Moreover, bio-inspired wax coatings, in particular composed of fluorinated waxes, demonstrated prominent antibiofouling properties achieved due to passive inhibition of bacterial attachment onto the coated surfaces.[27] Later, the coatings were found to be effective in biofilm establishment prevention over time in dairy food environment.[28] Other studies showed effective antifungal activity of the paraffin-based paper with incorporated essential oils.[29] Similar approach employing incorporation of active antimicrobials is used for gelatin-based films.[30] However, in these cases the antimicrobial effect is mainly attributed to the additives rather than to the matrix's material.

In the current study we aimed to form superhydrophobic coatings comprised of non-toxic saturated fatty acids (SFAs). These molecules are naturally present in biological systems, including human body, and are a part of human daily dietary uptake.[31–34] The latter makes fatty acids a promising candidate to serve as a coating agent for various applications with strict safety requirements, e.g. in biomedical field, food, pharma (including packaging), agriculture etc. In contrast to paraffin waxes, fatty acids contain a terminal carboxylic group, which may be involved in their antibacterial activity [35] and can, therefore, affect the coating's properties as well. Moreover, fatty acids are known as natural antimicrobial agents, which makes them even more advantageous for application in functional coatings due to the synergetic effect of their intrinsic superhydrophobic and antimicrobial properties.[35,36]

Several previous studies demonstrated the usage of SFAs for various coatings. An edible coating containing stearic acid was proposed for apple slices[37] and antibacterial activity of self-assembled palmitic and stearic acids on highly ordered pyrolytic graphite was proved.[32] Application of $SiO_2$ particles pre-coated with long carbon chain fatty acids on the surface was also suggested to form superhydrophobic coatings.[39] However, single-component SFA coatings were not a focus of these studies as well as the deposition techniques vary from those used in the current study.

Herein, we present facile deposition methods to form non-toxic superhydrophobic coatings on various types of surfaces using various SFAs. Resulted coatings were characterized in order to investigate their physical, crystallographic, structural and thermal stability properties. We also demonstrate the effect of the molecular length of SFAs and the effect of different deposition techniques on the coating's properties. In addition, the anti-biofouling and antimicrobial activity of the spray-deposited coatings was characterized against *Escherichia coli* and *Listeria innocua*, as relevant model bacteria.

**Experimental section**

*Sample preparation*:

I. Thermal deposition of SFAs on glass substrates was performed using Moorfield Minilab coating system. The process was performed in a vacuum chamber at a pressure of ~2×10$^{-9}$ [bar] by heating a crucible, which contains the coating material. Gradually increasing electrical current was applied in order to heat the crucible. The substrates were placed onto a rotating holder ~10 cm above the crucible. After deposition the samples were stored in a freezer (-25°C). An amount of 125±1 mg of a fatty acid was used for the deposition, unless is specified differently.

The following saturated fatty acids were used as coating agents: Palmitic acid (98%, Acros Organics, Malaysia), Stearic acid (98.5%, Sigma, Switzerland), Arachidic acid (99%, Sigma-Aldrich, India), Behenic acid (96%, AA Blocks, USA), Lignoceric acid (96%, AA Blocks, USA) and Cerotic acid (97%, AA Blocks, USA).

II. Spray coating was performed using commercially available dye spray gun, connected to an air compressor. The same system setup was used to perform the deposition of all fatty acids. Palmitic acid (98%, Acros Organics, Malaysia), stearic acid (97%, Merck, Germany) and arachidic acid (99%, AA-Blocks, USA) were used for the coatings` preparation. Ethanol (ABS AR, Gadot, Israel), acetone (AR, Bio-Lab,

Israel) or diethyl ether (stab.BHT, Bio-Lab, Israel) were used as solvents to prepare 20 mg/ml solutions. After the coatings were deposited, the samples were left overnight in vacuum oven at room temperature (RT) in order to remove any solvent residuals.

Thermal treatments of the coatings were performed using a Jeio Tech OV-11 oven at 50°C. The oven was pre-heated and samples were inserted once the temperature is stable. The samples were heated for 24 h and after cooling to RT placed into the freezer (-25°C) for storage.

*Characterization*:

Morphology of the coatings was studied using high-resolution scanning electron microscope (HR-SEM) Zeiss Ultra Plus FEG-SEM. Prior the imaging, a conductive carbon coating was deposited onto the surface of the samples. The same technique was used to image the cross-sections of the coatings. Prior to the coating deposition a scratch was implemented on the glass surface. Followed the deposition, samples were broken along the scratch and the exposed cross-sectional surface was observed using the HR-SEM.

Wetting properties of the coatings were characterized by CA and CAH measurements, which were performed using an Attension Theta Lite tensiometer and high-purity water or ethylene glycol (99.5%, Merck, Germany) droplets of 7 µL volume.

Roughness and coating thickness were measured using a dynamic confocal microscope (Leica DCM3D); data processing was performed using SensoMap Turbo software. The coating thickness was calculated as the difference in height between the lower and upper levels of a confocal profile measured after a scratch implementation on the coating using a 25G needle.

Structural characterization was performed using XRD measurements in a parallel beam theta-2theta mode using Cu anode sealed tube (Rigaku, SmartLab, X-ray

Diffractometer). The preferred orientation degree, $\eta$, was calculated according to the March-Dollase method[40] and it`s extension[41]: $\eta = [(1-r)^3/(1-r^3)]^{0.5} \cdot 100\%$. Parameter $r$ is dependent on the angle, $\alpha$, between the two compared planes, the preferred orientation plane and the reference plane and can be calculated as follows: $r = [sin^2\alpha/((k_{sample}/k_{powder})^{2/3} - cos^2\alpha)]^{1/3}$, where $k$ is the ratio between intensities of the preferred orientation plane and the reference plane (calculated for the sample and randomly oriented powder).

Differential scanning calorimetry (DSC) was used to study the origin of the additional XRD diffraction peaks. Fatty acids were detached from the glass substrate of spray-coated samples and were examined using DSC (LabSys 131 (SETARAM)). A cyclic measurement was performed by heating the powder from 20°C to 150°C followed by cooling down to 20°C.

*Bacterial cultures*:

Gram-negative *Escherichia coli* (*E. coli*) ATCC 8739 was cultured in Luria Broth (LB) medium containing 10 g L$^{-1}$ Bacto Tryptone (BD, USA), 5 g L$^{-1}$ Bacto yeast extract (BD, USA) and 5 g L$^{-1}$ sodium chloride (BioLab, Israel). LB agar plates for culturing were prepared by adding 18 g L$^{-1}$ Bacto agar (BD, USA) to the LB medium.

Gram-positive *Listeria innocua* (*L. innocua*) *ATCC 33090* was cultured in Brain Herat (BH) medium containing 37 g L$^{-1}$ BH Infusion (BD, USA). BH agar plates for culturing were prepared by adding 18 g L$^{-1}$ Bacto agar to the BH medium.

The bacteria were cultured in the appropriate agar plate and stored at 4 °C. Next, one bacteria colony was incubated overnight in 4 mL liquid medium (LB or BH) at 37°C under agitation (150 rpm) until the bacteria reached a stationary phase (~$10^9$ CFU mL$^{-1}$). Then, the bacterial suspensions were diluted by 1:100 in liquid medium for further experiments.

*Characterization of bacterial adhesion onto SFA spray-coated surfaces*:

SFA spray-coated surfaces were prepared as described in "Samples preparation" section using 10 mm round cover glass for slides as a substrate. The coated samples were UV sterilized prior to use. The samples placed into 6-well plates and incubated with 4 mL of the respective bacterial suspension at 37°C for 48±2 h. The samples were removed from the suspension and bacteria viability was studied by using a live/dead BacLight viability kit, where a 0.3% solution concentration (0.15% concentration of each reagent) was used. Subsequently, for three-dimensional image projection of the samples a confocal laser scanning microscope (CLSM), Zeiss LSM 510 META, was used. Combinations of 488 nm and 561 nm laser lines were used for the excitation of live bacteria and dead bacteria, respectively. Quantitation of adhered live/dead bacteria on the surface based on CLSM fluorescent images was performed using Spots analysis in Imaris 9.3.1 software. The values were normalized per depth unit to neutralize thickness difference of the coatings.

HR-SEM micrographs were obtained after bacteria were fixed on the surfaces using a glutaraldehyde solution (2% in 0.1 M normal saline) followed by dehydration through an ethanol series. Then, the samples were dried under vacuum overnight and sputtered with a conductive carbon coating.

*Antimicrobial studies*:

The antimicrobial properties of the different SFA powders were evaluated by in liquid medium via the drop-plate method. One bacteria colony (*E. coli* or *L. innocua*) was incubated overnight in 4 mL Nutrient Broth (NB) liquid medium (Sigma Aldrich, Israel) at 37°C under agitation (150 rpm) until the bacteria reached a stationary phase ($\sim 10^9$ CFU/mL). Then, the overnight bacteria culture was diluted in fresh NB medium ($\sim 10^7$ CFU/mL) and incubated for an additional 2h to achieve a logarithmic culture. Next, the logarithmic culture was diluted to $10^4$ CFU mL$^{-1}$ in 1:100 NB medium. 1 mL of the diluted bacterial suspension was incubated with 50 mg of different SFA powders at 37°C for 24 h in 24-well-plates under agitation (100 rpm). The cultures were decimally diluted and 10 µL drops were transferred onto NB solid agar substrate. The colonies were counted after 24 h incubation at 37°C.

**Results and Discussion**

Based on our previous studies on thermally deposited paraffin wax coatings, we, firstly, studied the feasibility of forming fatty acids coatings via thermal deposition method. To this end, a series of six SFAs was selected: palmitic acid (16C) – 16 carbons, stearic acid (18C) – 18 carbons, arachidic acid (20C) – 20 carbons, behenic acid (22C) – 22 carbons, lignoceric acid (24C) – 24 carbons and cerotic acid (26C) – 26 carbons. Selected SFAs were thermally deposited on glass microscope slides (see Experimental section) using an identical amount of 125±1 mg. In order to study the thermal stability of the SFA coatings, additional series of equivalent coatings were heated at 50° for 24h C in air. The properties of both as deposited and post-heated coatings were further studied as a function of the carbon chain length of selected SFAs.

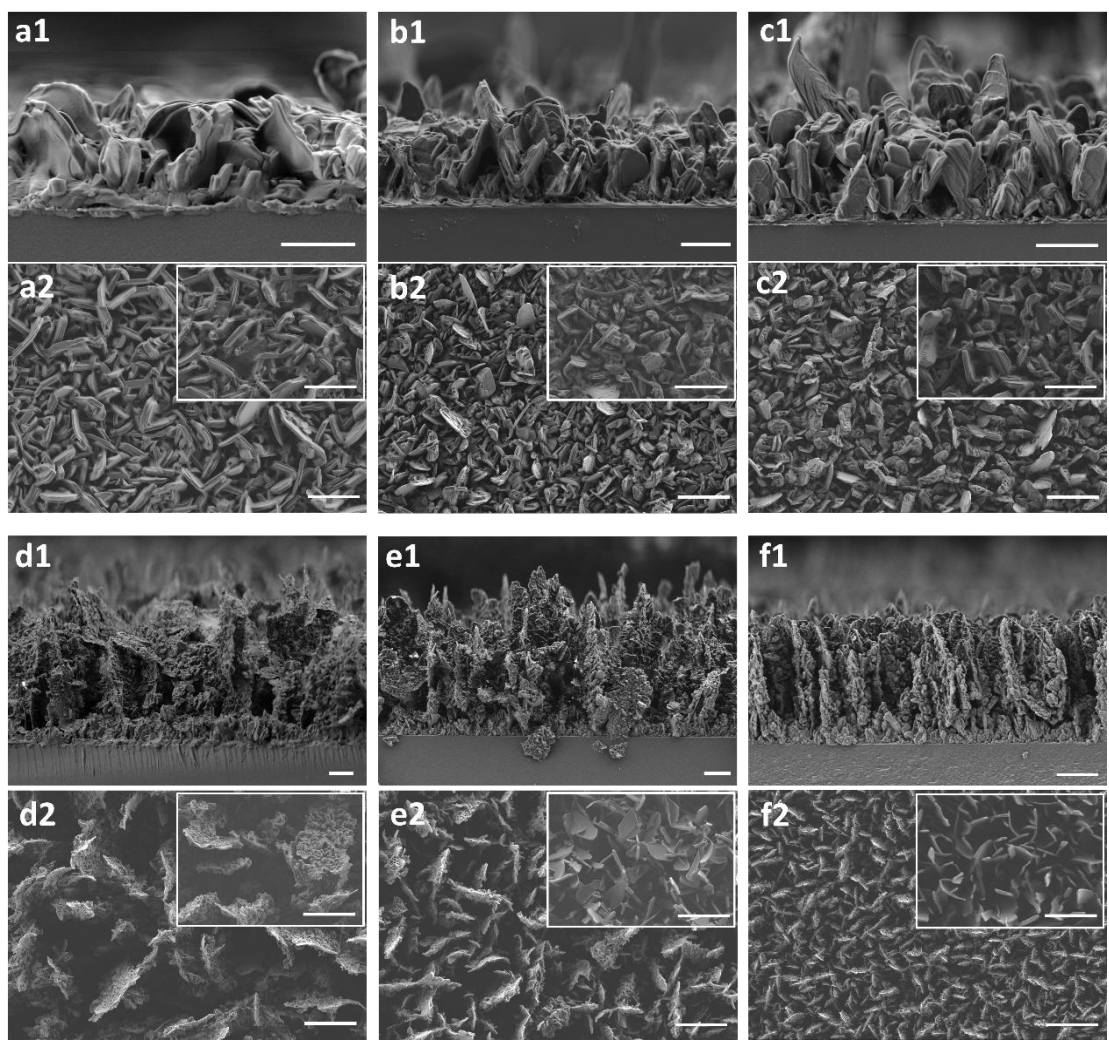

Figure 1: HR-SEM images of the deposited coatings before and after heating. Cross-sectional views of the coatings: a1) palmitic acid (16C), b1) stearic acid (18C), c1) arachidic acid (20C), d1) behenic acid (22C), e1) lignoceric acid (24C), f1) cerotic acid (26C). Scale bar is 2 μm. a2-f2) Planar views of the as deposited coatings, respectively. Scale bar is 4 μm. Insets – planar view of the coatings treated thermally for 24h at 50°C, respectively. Scale bar is 4μm.

The morphology of the resulted coatings was observed in cross-sectional and planar views acquired using a HR-SEM. As can be seen in Figure 1, all the SFA coatings appear as a dense uniform assembly of crystals with a characteristic morphology. Moreover, their surface morphology imaged via planar views (Figure 1 a1-f1) is also recognized across the whole depth of the coatings observed in corresponding cross-sectional views (Figure 1 a2-f2). Following a detailed analysis, we could distinguish two types of coatings: group A - SFAs with 16-20 carbons, including palmitic, stearic and arachidic acids; and group B – SFAs with 22-26 carbons, including behenic, lignoceric and cerotic acids. The coatings from the group A comprise well-defined and edged crystals ~2 μm in length and ~0.5 μm in thickness (Figure 1 a2-c2), while the crystals in the group B have smoother shape, showing a clear perpendicular orientation relative to the substrate. The backbone large crystals (2-6 μm) in group B are covered with smaller crystals up to few tens of nanometers in size (Figure 1 d2-f2), thereby developing a hierarchical structure. In the group A of SFAs the crystals' size is not significantly affected by the carbon chain length of fatty acids, which is proportional to the number of carbons in the molecule, (Figure 1 a2-c2). In contrast, the coatings in the group B show a clear correlation between the size of the crystals and the carbon chain length of a fatty acid. The longer the fatty acid, the smaller the obtained crystals: behenic acid (22C) crystals are ~6μm in length, lignoceric acid (24C) crystals are ~4μm in length and cerotic acid crystals are ~2μm in length (Figure 1 d2-f2). It can also be seen in the cross-section views, that the crystals of cerotic acid (26C) are better ordered than those of lignoceric acid (24C) and behenic acid (26C); this effect is probably achieved due to smaller size of cerotic acid (26C) crystals, which form denser and thinner coating, therefore less degrees of freedom in growth orientation exist.

The insets in Figure 1 a2-f2 show the morphology of each coating after heat treatment. While no significant change could be noticed in the case of the most of the fatty acid coatings, the coatings formed by lignoceric acid (24C) and cerotic acid (26C) underwent significant morphological transformation. In particular, the nanometric crystals, covering the backbone crystals prior to the heat treatment, disappeared (Fig. 1 e2-f2, insets). Interestingly, the morphology was changed in the case of the fatty acids with the high melting points of 75-88°C compared to melting points of unchanged coatings of SFAs – 63-80°C (SI 1: Common names and physical properties of the tested saturated fatty acids. ), even though they were expected to be more thermally stable at a given temperature.

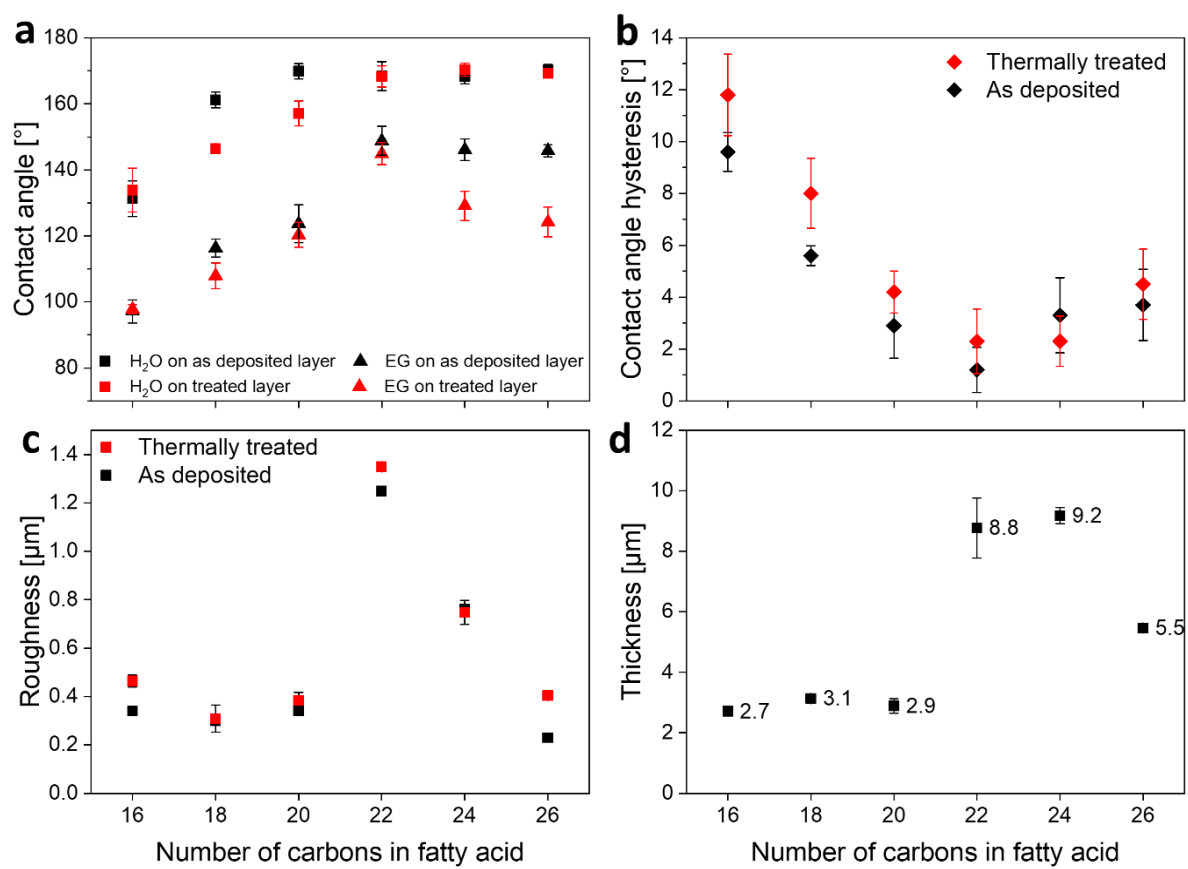

Figure 2: Characteristic properties of deposited fatty acid coatings as a function of carbons number in the molecule. Thermally treated coatings are presented by the red

color. a) CA of water and ethylene glycol on the as-deposited coatings (black squares and triangles) and on the thermally treated coatings (red squares and triangles). b) Water CAH on the as-deposited coatings (black rhombi) and thermally treated coatings (red rhombi). c) Roughness of as-deposited coatings (black squares) and thermally treated coatings (red squares). d) Coating thickness of as-deposited coatings.

Wetting properties of the coatings before and after heating were studied via CA and CAH measurements (Figure 2, a,b). As can be seen in Figure 2a, as-deposited coatings composed of SFAs with 18-26 carbons demonstrated superhydrophobic behavior resulting in water (high surface tension[42] – 72.8 mN/m$^2$) CA higher than 150⁰. Water CA increases as the carbon chain length extends from 16 to 20 carbons up to ~170° and remains stable for longer molecules from group B. Such behavior is similar for both as-deposited and heat-treated samples. Similar trend is seen in the case of ethylene glycol (lower surface tension[42] – 48 mN/m$^2$), however, its contact angle on post-heated lignoceric acid (24C) and cerotic acid (26C) decreases (Figure 2, a, red triangles), probably as a result of the change in crystal's morphology observed after heat treatment (Figure 1, a2-f2, insets). CAH results corroborate the CA values measured as a function of the number of carbons in a SFA (Figure 2, b): the higher the CA, the lower the CAH. Typically, low CAH values can be achieved on hierarchical superhydrophobic surfaces with good surface uniformity.[43,44,45] Therefore, the opposite trend of CAH values relatively to CA values of different coatings, combined with their morphology and the developed surface hierarchy of each coating was expected. The increase in CAH values in the case of the most samples after heating may indicate a higher surface heterogeneity caused by heat-induced crystals coarsening (Figure 1, a2-c2 insets).

The roughness and thickness of the studied coatings were measured by means of confocal microscope. The coatings from the group A with 16-20 carbons demonstrated similar roughness values ranging between 0.31-0.34 µm (Figure 2, c) as well as similar thicknesses of ~3 μm (Figure 2, d). In the case of the coatings from the group B, the highest roughness of ~1.3 μm is observed in the case of behenic acid (22C) coating, which can be explained by the bigger size of the crystals as seen in HR-SEM (see Figure 1, d2). Lignoceric acid (24C) and cerotic acid (26C) demonstrate a roughness of ~0.8 μm and ~0.2 μm, respectively. However, behenic acid (22C) and lignoceric acid (24C) coatings are of similar thicknesses of ~9 μm, while cerotic acid (26C) coating was measured to be ~6 μm thick, which is a third less than that of the SFAs in group B. The relatively low values of roughness and thickness can be ascribed to a smaller size and denser packing of the crystals comprising cerotic acid (26C) coatings as was observed in HR-SEM images (Fig.1, f1, f2). We also note that, thermal treatment had no significant effect on the roughness of all the SFA coatings (Figure 2c, red squares).

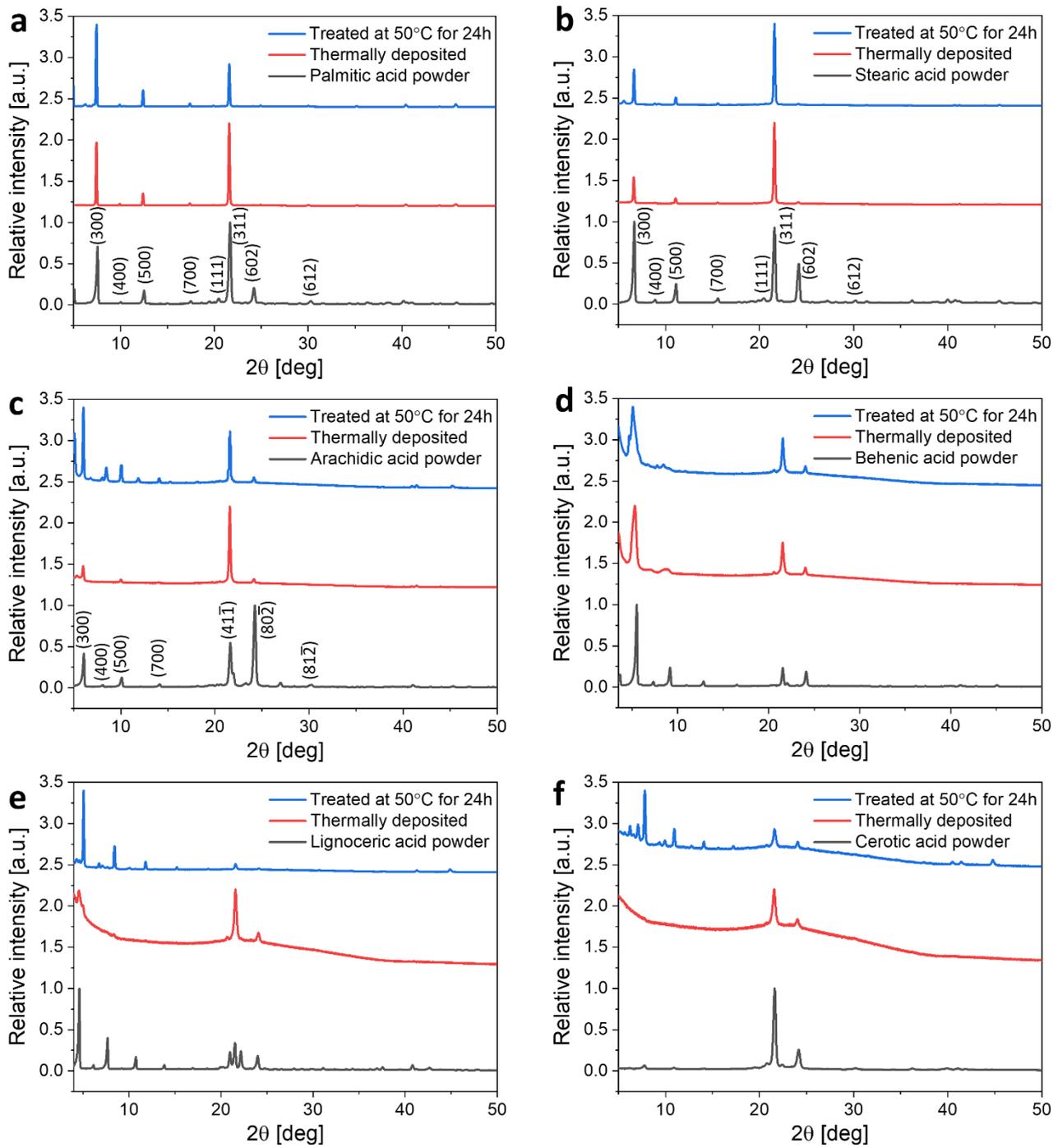

Figure 3: X-ray diffraction patterns of powdered fatty acids (black lines), as-deposited coatings (red lines) and thermally treated coatings (blue lines). a. Palmitic acid (16C), b. Stearic acid (18C), c. Arachidic acid (20C), d. Behenic acid (22C), e. Lignoceric acid (24C), f. Cerotic acid (26C).

In order to determine the crystal structure and crystallographic orientation of the as-deposited and heated coatings we collected X-ray diffraction patterns from the coatings and corresponding SFAs in a powdered form (Figure 3). The diffraction patterns collected from the coatings of the group A were indexed according to the existing literature.[46] As evidenced by the diffraction patterns, SFA coatings included into the group A are highly crystalline with a clear preferred crystal orientation (Figure 3, a-c, red lines). Palmitic acid (16C) and stearic acid (18C) coatings (Figure 3, a-b) demonstrate a degree of $(311)$ plane preferred orientation with $\eta$ (preferred orientation degree) of 47% and 45%, respectively, which is calculated relative to the $(602)$ plane (for details see Experimental section). Obtained rather low $\eta$ values indicate the presence of a high preferred orientation since the angle between the mentioned above planes is relatively high (55.99° and 55.92°, respectively). In the case of arachidic acid (20C) coating, a $(41\bar{1})$ preferred orientation is clearly seen (Figure 3, c). Analysis relative to the $(80\bar{2})$ plane of its powdered sample results in $\eta$ of 43% with an angle of 55.93° between the corresponding planes. The preferred orientation of the post-heated coatings was well maintained; minor changes in relative intensities of the diffraction peaks were observed after heat treatment of the coatings from the group A (Figure 3, a-c, blue lines). Additional diffraction peaks of the {h00} family emerged only in the case of the heat-treated arachidic acid (20C) coating (Figure 3, c).

The crystallinity of the SFA coatings from the group B decreased with an increase in the number of carbons of the fatty acids (Figure 3, d-f). A characteristic amorphous hump is most prominent in the diffraction pattern of the cerotic acid (26C) coating. We assume that, the longer the SFA molecule, the higher the diffusional barrier for crystallization, therefore, more material is deposited in the amorphous state. The diffraction patterns of the group B were not indexed since no crystallographic data is available for these SFAs. However, a clear preferred orientation can be identified from the difference in relative intensities of the corresponding diffraction peaks collected from the coatings and powdered form of SFAs (Figure 3, a-f). Heat treatment induced structural reorganization of the coatings (Figure 3, d-f, blue lines). Improvement in their crystallinity resulted in the reducing of the amorphous hump and appearance of additional diffraction peaks (Figure 3, d-f, blue lines).

Additional investigation was performed in the case of behenic acid (22C) coating, where the deposition was applied on a substrate pre-heated to $45\pm2$ °C in order to reduce the diffusional barrier during the crystal formation. These results proved that additional thermal energy given at the pre-deposition stage strongly affects the morphology, crystallinity and the roughness of the resulted coating (for additional information see SI 2).

We also studied the influence of the amount of the deposited SFAs on the properties of the obtained coatings. The experimental data fully support our previous findings: higher amount of the deposited of SFA of the group A caused crystals coarsening, while in the group B it led to the development of more prominent hierarchical structure, followed by change in the coatings` properties (for additional information see **Error! Reference source not found.**).

Considering the limited application and scalability potential of the thermal deposition method, we further aimed at developing and studying a more facile and practical approach to form superhydrophobic SFA coatings utilizing a spraying technique. In contrast to thermal deposition, spraying offers a great simplicity and high versatility enabling fast application, while only basic facilities could be required.

The first challenge was to define a suitable solvent allowing fabrication of the SFA-based spray solutions. We, therefore, studied the use of the three different solvents, namely ethanol, acetone and diethyl ether, in a test case of stearic acid (18C)-based spray. Resulted solutions were spray-deposited onto glass substrates and obtained coatings were characterized as for their morphology, wetting properties and surface roughness (Figure 4).

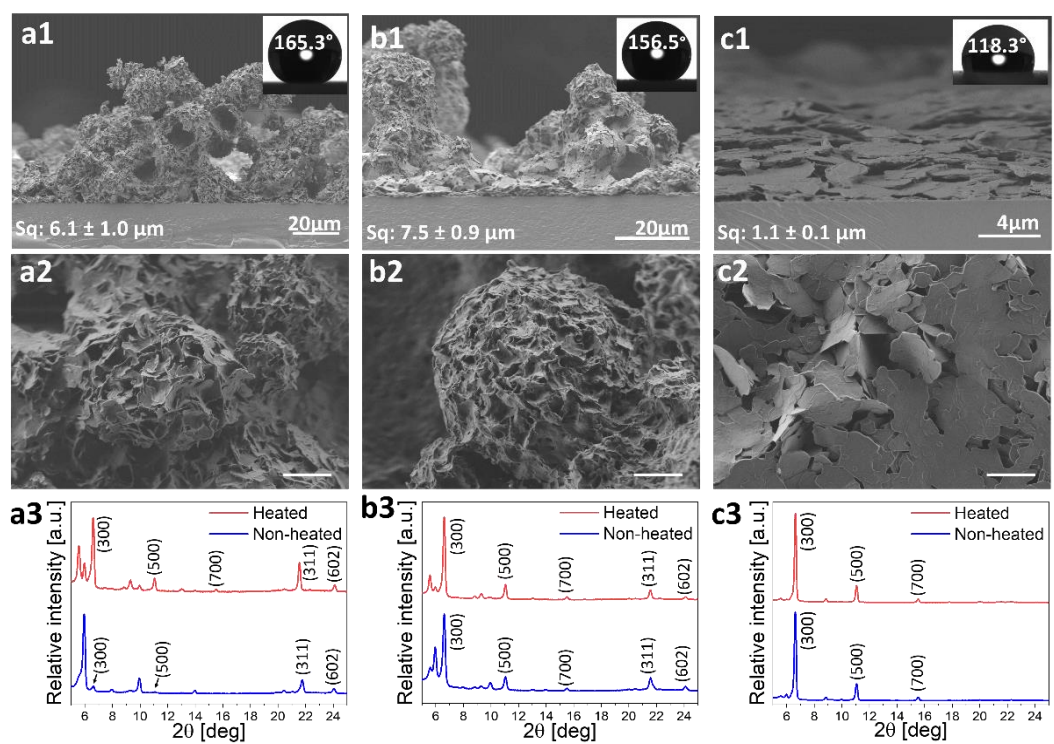

Figure 4: HR-SEM and XRD characterization of stearic acid (18C) spray coatings based on various solvents: a) diethyl ether, b) acetone, c) ethanol. a1-c1) cross-sectional views, respectively. Insets: water CA and roughness values. a2-c2) planar views of spray coatings, respectively. Scale bar is 4μ. a3-c3) XRD of spray coatings before and after heating, respectively (blue and red lines respectively).

Planar and cross-sectional imaging of the spray-deposited coatings revealed that in the case of diethyl ether- and acetone-based stearic acid (18C) coatings the surface is covered with sphere-shaped aggregates that are tens of microns in size and composed of smaller micron-sized crystals (Figure 4 a1 – b1 and a2 – b2). Such hierarchical surface structure resulted in high roughness values of 6.1±1 and 7.5±0.9 μm, respectively. Both coatings demonstrated superhydrophobic behavior with water CA higher than 156º and CAH lower than 7º (see insets in Figure 4, a1-b1). Reasonably poorer superhydrophobic performance with a CA of 118.3±5.8°and a CAH of 11º was observed in the case of ethanol-based stearic acid (18C) spray coating, whose surface comprises closely packed platelet-like crystals with smooth surface (Figure 4, c1-c2). As no hierarchical structure was formed, a rather low, as compared to those of acetone-based and diethyl ether-based coatings, surface roughness of 1.1±0.1 µm was measured. Based on the obtained results, we could assume that the properties of the spray coatings depend on the volatility of the solvent. As the solvent is more volatile (in our case $T_{b, ethanol}$=78°C > $T_{b, acetone}$=56°C > $T_{b, diethyl ether}$=34°C [47]) it evaporates faster so as facilitates the formation of SFA crystals aggregates and their hierarchical organization, which in turns results in the improvement of the coatings' wetting properties. Among the three chosen solvents, diethyl ether-based stearic acid (18C) coating demonstrated the most optimal superhydrophobic properties.

The structure of the spray coatings was studied via XRD (Figure 4 a3 – c3, blue lines). A strong preferred orientation along the {h00} planes was observed in the case of stearic acid spray coating based on the EtOH solution as well as acetone solution (Figure 4, a3-b3), as compared to the structure of its powdered form (Figure 3, b). The diffraction pattern collected from the diethyl ether-based coating (Figure 4 c3) also differs from that of randomly-oriented stearic acid powder (Figure 3, b) revealing the presence of preferred orientation, however, the {h00} planes are missing at their expected *2θ* values. Interestingly, unexpected diffraction peaks emerged in the case of acetone-based coating as well as diethyl-based coating (Figure 4 c1-c2). These diffraction peaks are left-shifted relatively to (300) and (500) planes. Since the solvents are volatile and evaporate extremely fast when sprayed, new diffraction peaks may appear due to change in *d*-spacings caused by residual strains in the crystal lattice. The calculation of such residual strains results in a value of 10.8-11.0%, which seems reasonable since the strain is distributed along the carbon chain of the molecule. In order to facilitate relaxation of strains, the coatings were heated for 24h at 50°C and characterized again by XRD (Figure 4 c1-c3, red lines). The XRD of the heated coatings indeed showed that relative intensity of the additional peak decreased, while (300) plane intensity increased. This finding together with DCS analysis (SI 4) support the assumption of the presence of strains in coatings' lattice due to extremely fast solvent evaporation.

We further used diethyl ether solvent in order to fabricate palmitic acid (16C) and arachidic acid (20C) spray coatings, since these fatty acids showed a good thermal stability when were deposited via thermal deposition method. As expected, the morphology, wetting properties, roughness, and crystallographic patterns of the obtained coatings were similar to those of the stearic acid coating deposited using the same solvent (for additional information see SI 7).

As compared to the corresponding thermally deposited coatings (Fig. 1, a-c), sprayed coatings surfaces are more heterogeneous with roughness values more than one order of magnitude larger than those of thermally deposited coatings (Figure 2 c, Figure 4 a1, SI 7). Additionally, prominent hierarchical morphology improves the superhydrophobic properties of the sprayed coatings. Indeed, the palmitic acid (16C) sprayed coating, for example, exhibited a CA>150°, while its thermally deposited counterpart demonstrated a CA lower than 140° (see SI 5). Overall, based on our observations, spray coating method was found to be easier in its application and provided better superhydrophobic characteristics to the coatings.

Our previous studies demonstrated that in addition to the unique physical, wetting, and crystallographic properties of self-assembled wax coatings, they also exhibit exceptional anti-biofouling properties achieved via passive inhibition of bacterial adhesion to the surface.[27,28] The anti-biofouling and potential antimicrobial properties of the SFA coatings, formed via spraying, were characterized using two common model bacteria, *E. coli* and *L. innocua*. The latter is a Gram-positive bacterium and a well-known indicator for the pathogenic *Listeria monocytogenes*.[48] While, *E. coli* is a Gram-negative bacterium, which is usually used as a model indicator bacterium since it is a typical inhabitant of the human intestinal tract.[49]

Bacterial attachment and viability onto the SFA coatings were characterized by live/dead staining followed by CLSM imaging analysis. **Error! Reference source not found.** a-b shows representative three-dimensional orthogonal projection images of the coatings, depicting stained adhered bacterial cells where live/dead cells appear in green and red, respectively. Qualitatively, it is apparent for both species that the total number of adhered cells on all coated surfaces is reduced compared to the control uncoated substrate. For a more quantitative assessment, we used image analysis to calculate the relative bacterial cells surface density, termed as RCD (i.e., the number of cells per unit area, normalized per unit depth, presented relatively to control uncoated surface), and the respective values for *E. coli* are depicted in **Error! Reference source not found.** a1-a4 insets. Note that in the case of *L. innocua* (**Error! Reference source not found.** b1-b4), due to the dense bacteria population such analysis could not be performed as individual cells could not be easily distinguished. For palmitic acid (16C) coatings a small reduction in adherent cells is observed and the RCD value is decreased by 20% (**Error! Reference source not found.**, a1 vs. a2, b1 vs. b2). Stearic (18C) and arachidic (20C) acid coatings exhibit a profound anti-biofouling effect against both species, and *E. coli* adhesion is reduced by ~90-93% (Figure 5 a1 vs. a3 and a4, b1 vs. b3 and b4). Interestingly, in the case of stearic and arachidic acids no *L. innocua* cells were detected by means of CLSM imaging, see Figure 5 b3-4.

In addition to the reduction of adhered cells, the proportion of live cells decreases for all surfaces, indicating a biocidal effect of the coating, particularly in the case of palmitic acid (16C) (Figure 5 a1 vs. a2, b1 vs. b2 insets). Yet, it is important to note that in the case of stearic (18C) and arachidic (20C) acid coatings their strong anti-biofouling effect, resulting in significant reduction in bacteria count, may interfere with the assessment of their biocidal action (Figure 5 a1 vs. a3-a4). This is further manifested in the case of *L. innocua* on stearic (18C) and arachidic (20C) acid coatings, where no fluorescence signal was detected in CSLM imaging.

Based on the CLSM studies we can deduce that there are two modes of action by which SFA surfaces function: anti-biofouling and biocidal. Such dual mode mechanism of action is reported in the literature for both synthetic[50] and natural superhydrophobic surfaces[51]. In order to elucidate the role of each of these effects, we studied the intrinsic antimicrobial activity of the respective fatty acids when in powdered form (within bacteria media) and the results are summarized in Table SI 8. No inhibition effect of the powdered fatty acids was detected against *E. coli*; yet, for *L. innocua*, growth reduction of one order of magnitude was obtained for powdered palmitic (16C) and stearic (18C) acids (see Table in SI 8). Surprisingly, powdered arachidic acid (20C) completely inhibited the growth of *L. innocua*. The intrinsic antibacterial properties of fatty acids are well documented, and unsaturated fatty acids are reported to exhibit superior potency against Gram-positive bacteria.[35,36,52,53] Yet, their exact mode of action is not fully understood, but may include disruption of membrane and electron transport chain of bacteria cells, inhibition of enzymes activity and protein synthesis. Thus, we suggest that the observed antibacterial properties of these coatings can be ascribed to the unique combination of their surface structure, which can induce cells rupture due to stretching and puncturing of the cell membrane,[54,55] as well as the specific intrinsic biocidal activity of the used fatty acid. Therefore, the balance between these two effects may be contextual and species specific.

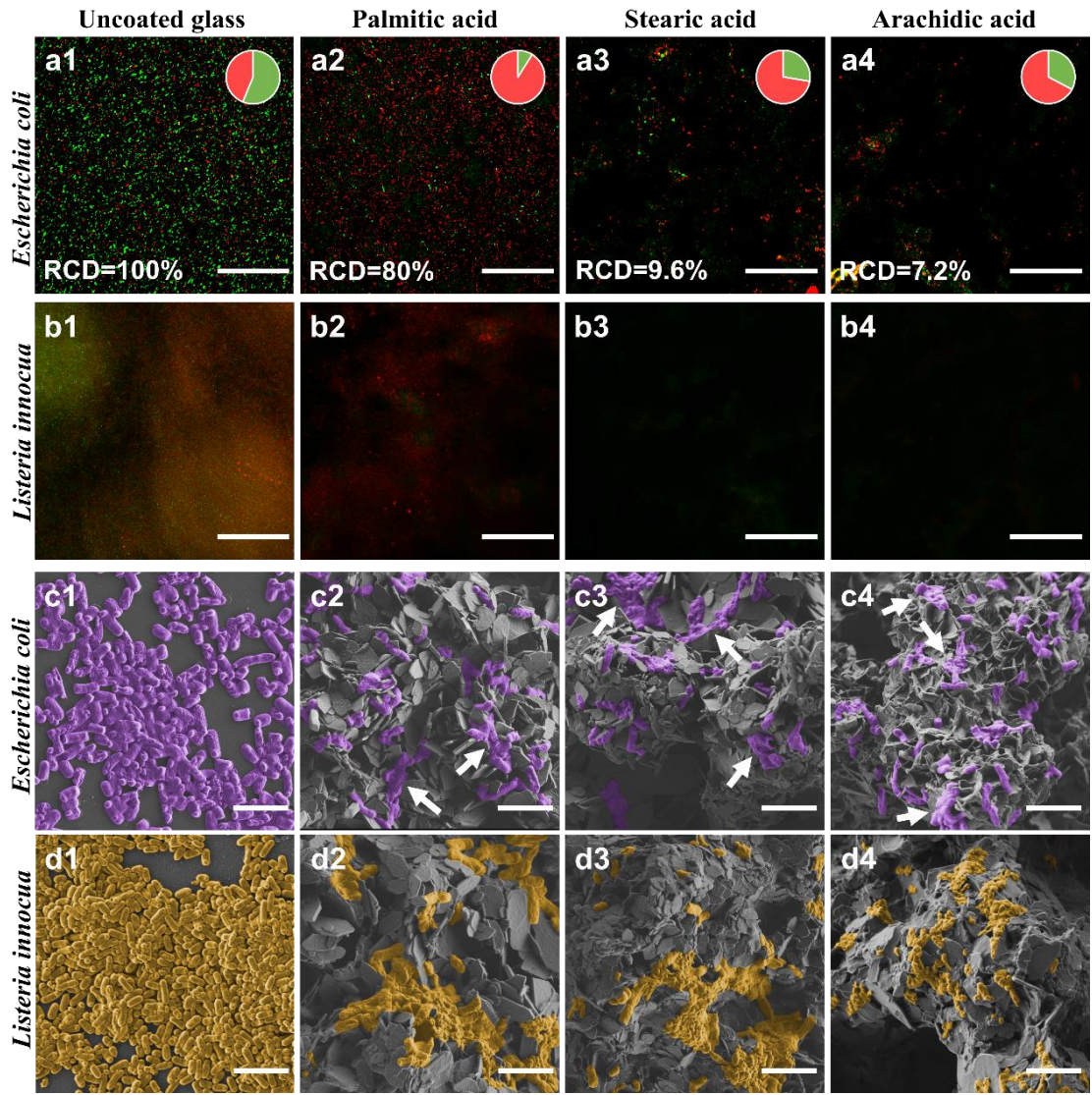

Figure 5: CLSM orthogonal projection images of adhered (a) E. coli and (b) L. innocua to spray-coated SFA surfaces: 1. Uncoated glass (control), 2. Palmitic acid (16C), 3. Stearic acid (18C), 4. Arachidic acid (20C). Scale bar is 80 μm. Insets: relative bacteria cells density (RCD) per unit area, normalized per unit depth (cells μm$^{-3}$); green and red sections represent live and dead cells, respectively. HR-SEM images of adhered (c) E. coli and (d) L. innocua to the sprayed SFA surfaces: 1. Uncoated glass (control), 2. Palmitic acid (16C), 3. Stearic acid (18C), 4. Arachidic acid (20C). Scale bar is 4 μm. Bacteria cells are false-colored to ease observation.

Next, we used HR-SEM as a complementary tool to qualitatively assess bacteria adhesion to the coatings. Given the highly hierarchical morphology of the coatings and their complex architecture (Figure 4 a, SI 7 a, b), bacteria may not be fully exposed to the microscope laser beam using CLSM technique due to their adhesion to hidden locations on the surface and could result in their lower detection. Moreover, surface structure is one of the factors that may strongly affect the bacteria adhesion and their behavior on the surfaces.[56–58] Similarly to the CLSM results, bacterial cells density on the SFA coatings is profoundly reduced in comparison to the uncoated substrates (**Error! Reference source not found.** c1 vs. c2-c4 and d1 vs. d2-d4) which is attributed to the anti-biofouling effect of the coatings. Moreover, the electron micrographs reveal that the morphology of the SFA-adherent cells is altered in comparison to their characteristic appearance, observed on the uncoated substrate. Such morphological changes of the adhered cells are indicative of damaged cells.[54,55,59,60] For *E. coli*, white arrows in **Error! Reference source not found.** c2-c4 indicate cells which appear to be severely damaged, supporting the CLSM findings of reduced cells viability on the SFA coatings. In the case of *L. innocua*, it appears that the majority of cells residing on the SFA coated surfaces have lost their integrity and are badly deformed and damaged (**Error! Reference source not found.** d1 vs. d2-d4). These morphological changes agree well with the CLSM results and may be ascribed to a combined antibacterial functionality of these coatings exerted by their hierarchical structure and intrinsic biocidal activity of the fatty acids (**Error! Reference source not found.** b1-b4). Nonetheless, the balance between these two effects is likely to dictate their specific activity against different species, as manifested by superior activity of arachidic acid (20C) coatings against *L. innocua* cells.

To summarize, spray-deposited SFA coatings demonstrate an effective prevention of *E. coli* and *L. innocua* adhesion over two days incubation accompanied with additional strong biocidal effect against *L. innocua*, and weak biocidal effect against *E. coli*.

**Conclusions**

In this study we developed novel superhydrophobic coatings comprised of SFAs. We demonstrated the feasibility of the thermal deposition approach as a synthetic route to form superhydrophobic coatings with low CAH (<10°) comprised of SFAs with various chain lengths. By varying the molecule's length and deposited weight of selected SFA it is possible to control and tune the properties of the obtained coatings such as their CA, CAH, thickness and surface roughness. Even though the thermal treatment induced structural and morphological changes of the coatings, we showed that SFA coatings maintain their superhydrophobic wetting properties after annealing up to 50°C for at least 24 h.

Moreover, we developed an alternative facile spray coating method established via SFA solvent-based formulations enabling a broad application of the coatings. The spray-coated surfaces demonstrated excellent superhydrophobicity, prominent hierarchical structure and highly preferred orientation of the crystals. Most importantly, our preliminary results show that these coatings exhibit a unique combination of anti-biofouling and antibacterial properties against *E. coli* and *L. innocua*, used as relevant model bacteria. We suggest that the specific activity of the coatings against different bacteria stems from their complex hierarchal structure and the fatty acid intrinsic biocidal effect. Thus, these coatings may be potentially tailored to exhibit a wide repertoire of desired antibacterial properties for different applications.

This new family of multifunctional coatings displays superior properties of superhydrophobicity as well as anti-biofouling and antimicrobial activity, while being safe and sustainable by design. We believe that this work provides a proof of concept for reliability of fatty acids to serve as functional antimicrobial coatings and further research may expand the range of properties that can be achieved for saturated fatty acid-based coatings.

## ASSOCIATED CONTENT

**Supporting Information**

SFA properties table; characterization of thermally deposited behenic acid on heated substrate (HR-SEM, CA, roughness, XRD); weight-dependent coatings characterization (HR-SEM, CA, roughness, XRD); sprayed SFA coatings characterization (HR-SEM, CA, roughness, XRD, DSC); bacteria count after

## AUTHOR INFORMATION

**Corresponding Authors**

**Boaz Pokroy** - Department of Materials Science and Engineering, Technion − Israel Institute of Technology, 32000 Haifa, Israel.

**Ester Segal** - Faculty of Biotechnology and Food Engineering, Technion − Israel Institute of Technology, 32000 Haifa, Israel

## ACKNOWLEDGEMENTS

We thank Dr. Inna Zeltser from laboratory for physical measurements, Department of Materials Science and Engineering, Technion - Israel Institute of Technology for assisting in DSC measurements. We are indebted to the Lokey Interdisciplinary Center for Life Sciences and Engineering, Technion - Israel Institute of Technology for helping in setting up the CLSM measurement and data processing.

## ABBREVIATIONS

CA – contact angle; CAH – contact angle hysteresis; SFA – saturated fatty acid

# REFERENCES

1.  Sun, J. & Bhushan, B. Hierarchical structure and mechanical properties of nacre: A review. *RSC Advances* (2012) doi:10.1039/c2ra20218b.

2.  Aizenberg, J., Tkachenko, A., Weiner, S., Addadi, L. & Hendler, G. Calcitic microlenses as part of the photoreceptor system in brittlestars. *Nature* (2001) doi:10.1038/35090573.

3.  Barthlott, W. & Neinhuis, C. Purity of the sacred lotus, or escape from contamination in biological surfaces. *Planta* (1997) doi:10.1007/s004250050096.

4.  Schüler, D. & Frankel, R. B. Bacterial magnetosomes: Microbiology, biomineralization and biotechnological applications. *Applied Microbiology and Biotechnology* (1999) doi:10.1007/s002530051547.

5.  Polishchuk, I. *et al.* Coherently aligned nanoparticles within a biogenic single crystal: A biological prestressing strategy. *Science (80-. ).* (2017) doi:10.1126/science.aaj2156.

6.  Barthlott, W., Mail, M. & Neinhuis, C. Superhydrophobic hierarchically structured surfaces in biology: Evolution, structural principles and biomimetic applications. *Philos. Trans. R. Soc. A Math. Phys. Eng. Sci.* (2016) doi:10.1098/rsta.2016.0191.

7.  Darmanin, T. & Guittard, F. Superhydrophobic and superoleophobic properties in nature. *Materials Today* (2015) doi:10.1016/j.mattod.2015.01.001.

8.  Rich, B. B. & Pokroy, B. A study on the wetting properties of broccoli leaf surfaces and their time dependent self-healing after mechanical damage. *Soft Matter* (2018) doi:10.1039/C8SM01115J.

9.  Young, T. An essay on cohesion of fluids. *Philos. Trans. te R. Soc. London* (1805).

10.     Wenzel, R. N. Resistance of solid surfaces to wetting by water. *Ind. Eng. Chem.* (1936) doi:10.1021/ie50320a024.

11.     Cassie, B. D., Cassie, A. B. D. & Baxter, S. Of porous surfaces,. *Trans. Faraday Soc.* (1944) doi:10.1039/tf9444000546.

12.     de Gennes, P.-G., Brochard-Wyart, F. & Quéré, D. *Capillarity and Wetting Phenomena*. *Capillarity and Wetting Phenomena* (2004). doi:10.1007/978-0-387-21656-0.

13.     Si, Y., Dong, Z. & Jiang, L. Bioinspired Designs of Superhydrophobic and Superhydrophilic Materials. *ACS Cent. Sci.* **4**, 1102–1112 (2018).

14.     Li, W., Zhan, Y. & Yu, S. Applications of superhydrophobic coatings in anti-icing: Theory, mechanisms, impact factors, challenges and perspectives. *Prog. Org. Coatings* **152**, 106117 (2021).

15.     Jokinen, V., Kankuri, E., Hoshian, S., Franssila, S. & Ras, R. H. A. Superhydrophobic Blood-Repellent Surfaces. *Adv. Mater.* **30**, (2018).

16.     Razavi, S. M. R. *et al.* Environment-Friendly Antibiofouling Superhydrophobic Coatings. *ACS Sustain. Chem. Eng.* **7**, 14509–14520 (2019).

17.     Han, Z., Feng, X., Guo, Z., Niu, S. & Ren, L. Flourishing Bioinspired Antifogging Materials with Superwettability: Progresses and Challenges. *Adv. Mater.* **30**, (2018).

18.     Han, Z. *et al.* Bio-inspired antifogging PDMS coupled micro-pillared superhydrophobic arrays and SiO2 coatings. *RSC Adv.* **8**, 26497–26505 (2018).

19.     Yan, Y. Y., Gao, N. & Barthlott, W. Mimicking natural superhydrophobic surfaces and grasping the wetting process: A review on recent progress in preparing superhydrophobic surfaces. *Advances in Colloid and Interface Science* (2011) doi:10.1016/j.cis.2011.08.005.

20.    Parvate, S., Dixit, P. & Chattopadhyay, S. Superhydrophobic Surfaces: Insights from Theory and Experiment. *J. Phys. Chem. B* **124**, 1323–1360 (2020).

21.    Jeevahan, J., Chandrasekaran, M., Britto Joseph, G., Durairaj, R. B. & Mageshwaran, G. Superhydrophobic surfaces: a review on fundamentals, applications, and challenges. *J. Coatings Technol. Res.* **15**, 231–250 (2018).

22.    Kumar, A. & Nanda, D. *Methods and fabrication techniques of superhydrophobic surfaces*. *Superhydrophobic Polymer Coatings: Fundamentals, Design, Fabrication, and Applications* (Elsevier Inc., 2019). doi:10.1016/B978-0-12-816671-0.00004-7.

23.    Liquid-Infused Nanostructured Surfaces with Extreme Anti-Ice and Anti-Frost Performance.

24.    Pechook, S. & Pokroy, B. Self-assembling, bioinspired wax crystalline surfaces with time-dependent wettability. *Adv. Funct. Mater.* (2012) doi:10.1002/adfm.201101721.

25.    Pechook, S., Kornblum, N. & Pokroy, B. Bio-inspired superoleophobic fluorinated wax crystalline surfaces. *Adv. Funct. Mater.* (2013) doi:10.1002/adfm.201203878.

26.    Pechook, S. & Pokroy, B. Bioinspired hierarchical superhydrophobic structures formed by n-paraffin waxes of varying chain lengths. *Soft Matter* (2013) doi:10.1039/c3sm27484e.

27.    Pechook, S. *et al.* Bioinspired passive anti-biofouling surfaces preventing biofilm formation. *J. Mater. Chem. B* (2015) doi:10.1039/c4tb01522c.

28.    Ostrov, I., Polishchuk, I., Shemesh, M. & Pokroy, B. Superhydrophobic Wax Coatings for Prevention of Biofilm Establishment in Dairy Food. *ACS Appl. Bio Mater.* (2019) doi:10.1021/acsabm.9b00674.

29.    Rodriguez-Lafuente, A., Nerin, C. & Batlle, R. Active paraffin-based paper packaging for extending the shelf life of cherry tomatoes. *J. Agric. Food Chem.* **58**, 6780–6786

(2010).

30. Ramos, M., Valdés, A., Beltrán, A. & Garrigós, M. Gelatin-Based Films and Coatings for Food Packaging Applications. *Coatings* **6**, 41 (2016).

31. Calder, P. C. Functional Roles of Fatty Acids and Their Effects on Human Health. *Journal of Parenteral and Enteral Nutrition* (2015) doi:10.1177/0148607115595980.

32. Tvrzicka, E., Kremmyda, L. S., Stankova, B. & Zak, A. Fatty acids as biocompounds: Their role in human metabolism, health and disease - a review. part 1: Classification, dietary sources and biological functions. *Biomedical Papers* (2011) doi:10.5507/bp.2011.038.

33. Agriculture, U. S. D. of H. and H. S. and U. S. D. of. 2015 − 2020 Dietary Guidelines for Americans. *2015 − 2020 Diet. Guidel. Am. (8th Ed.* (2015) doi:10.1097/NT.0b013e31826c50af.

34. Spector, A. A. Essentiality of fatty acids. *Lipids* (1999) doi:10.1007/bf02562220.

35. Desbois, A. P. & Smith, V. J. Antibacterial free fatty acids: Activities, mechanisms of action and biotechnological potential. *Applied Microbiology and Biotechnology* (2010) doi:10.1007/s00253-009-2355-3.

36. Casillas-Vargas, G. *et al.* Antibacterial fatty acids: An update of possible mechanisms of action and implications in the development of the next-generation of antibacterial agents. *Prog. Lipid Res.* **82**, (2021).

37. Chiumarelli, M. & Hubinger, M. D. Evaluation of edible films and coatings formulated with cassava starch, glycerol, carnauba wax and stearic acid. *Food Hydrocoll.* (2014) doi:10.1016/j.foodhyd.2013.11.013.

38. Ivanova, E. P. *et al.* Bactericidal activity of self-assembled palmitic and stearic fatty acid crystals on highly ordered pyrolytic graphite. *Acta Biomater.* (2017)

doi:10.1016/j.actbio.2017.07.004.

39.   Heale, F. L. *et al.* Inexpensive and non-toxic water repellent coatings comprising SiO2

      nanoparticles and long chain fatty acids. *RSC Adv.* **8**, 27064–27072 (2018).

40.   Dollase, W. A. CORRECTION OF INTENSITIES OF PREFERRED ORIENTATION IN

      POWDER DIFFRACTOMETRY: APPLICATION OF THE MARCH MODEL. *J. Appl.*

      *Crystallogr.* (1986) doi:10.1107/S0021889886089458.

41.   Zolotoyabko, E. Determination of the degree of preferred orientation within the

      March-Dollase approach. *J. Appl. Crystallogr.* (2009)

      doi:10.1107/S0021889809013727.

42.   van Oss, C. J. Acid-base interfacial interactions in aqueous media. *Colloids and*

      *Surfaces A: Physicochemical and Engineering Aspects* (1993) doi:10.1016/0927-

      7757(93)80308-2.

43.   Marmur, A., Volpe, C. Della, Siboni, S., Amirfazli, A. & Drelich, J. W. Contact angles

      and wettability: Towards common and accurate terminology. *Surf. Innov.* (2017)

      doi:10.1680/jsuin.17.00002.

44.   Yuan, Y. & Lee, T. R. Chapter 1: Contact Angle and Wetting Properties. in *Surface*

      *Science Techniques* (2013). doi:10.1007/978-3-642-34243-1.

45.   Rudawska, A. *Assessment of surface preparation for the bonding/adhesive*

      *technology*. *Surface Treatment in Bonding Technology* (2019). doi:10.1016/b978-0-

      12-817010-6.00009-6.

46.   Moreno-Calvo, E. *et al.* Competing intermolecular interactions in the high-

      temperature solid phases of even saturated carboxylic acids (C10H19O2H to

      C20H39O2H). *Chem. - A Eur. J.* **15**, 13141–13149 (2009).

47.   Prat, D. *et al.* CHEM21 selection guide of classical- and less classical-solvents. *Green*

*Chem.* (2015) doi:10.1039/c5gc01008j.

48.     Shemesh, R. *et al.* Novel LDPE/halloysite nanotube films with sustained carvacrol release for broad-spectrum antimicrobial activity. *RSC Adv.* **5**, 87108–87117 (2015).

49.     Blount, Z. D. The unexhausted potential of E. coli. *Elife* **4**, 1–12 (2015).

50.     Hu, J. *et al.* A new anti-biofilm strategy of enabling arbitrary surfaces of materials and devices with robust bacterial anti-adhesion: Via a spraying modified microsphere method. *J. Mater. Chem. A* **7**, 26039–26052 (2019).

51.     Elbourne, A., Crawford, R. J. & Ivanova, E. P. Nano-structured antimicrobial surfaces: From nature to synthetic analogues. *J. Colloid Interface Sci.* **508**, 603–616 (2017).

52.     McGaw, L. J., Jäger, A. K. & Van Staden, J. Antibacterial effects of fatty acids and related compounds from plants. *South African J. Bot.* **68**, 417–423 (2002).

53.     Yuyama, K. T., Rohde, M., Molinari, G., Stadler, M. & Abraham, W. R. Unsaturated fatty acids control biofilm formation of staphylococcus aureus and other gram-positive bacteria. *Antibiotics* **9**, 1–11 (2020).

54.     Fisher, L. E. *et al.* Bactericidal activity of biomimetic diamond nanocone surfaces. *Biointerphases* **11**, 011014 (2016).

55.     Hazell, G., Fisher, L. E., Murray, W. A., Nobbs, A. H. & Su, B. Bioinspired bactericidal surfaces with polymer nanocone arrays. *J. Colloid Interface Sci.* **528**, 389–399 (2018).

56.     Cheng, Y., Feng, G. & Moraru, C. I. Micro-and nanotopography sensitive bacterial attachment mechanisms: A review. *Front. Microbiol.* **10**, 1–17 (2019).

57.     Hsu, L. C., Fang, J., Borca-Tasciuc, D. A., Worobo, R. W. & Moraru, C. I. Effect of micro-and nanoscale topography on the adhesion of bacterial cells to solid surfaces. *Appl. Environ. Microbiol.* **79**, 2703–2712 (2013).

58.     Leonard, H. *et al.* Shining light in blind alleys: deciphering bacterial attachment in

silicon microstructures. *Nanoscale Horizons* **7**, (2022).

59.    Bermúdez-Aguirre, D., Mawson, R. & Barbosa-Cánovas, G. V. Study of possible

mechanisms of inactivation of Listeria innocua in thermo-sonicated milk using

scanning electron microscopy and transmission electron microscopy. *J. Food Process.*

*Preserv.* **35**, 767–777 (2011).

60.    Leonard, H. *et al.* Rapid diagnostic susceptibility testing of bacteria and fungi from

clinical samples using silicon gratings. **1089504**, 3 (2019).

**Supporting Information**

SI 1: Common names and physical properties of the tested saturated fatty acids.

| Fatty acid name | No. of carbons | Melting temperature [°C][32] |
|---|---|---|
| Palmitic acid | 16 | $62.5 - 63.1$ |
| Stearic acid | 18 | $67 - 69.6$ |
| Arachidic acid | 20 | $75.3 - 75.4$ |
| Behenic acid | 22 | $79.9 - 80.0$ |
| Lignoceric acid | 24 | $75 - 83$ |
| Cerotic acid | 26 | $87 - 88$ |

In order to investigate the significant morphological difference of the coatings from group A and group B, deposition of behenic acid (22C) was performed during a substrate heating to 45±2°C to reduce the diffusional barrier (

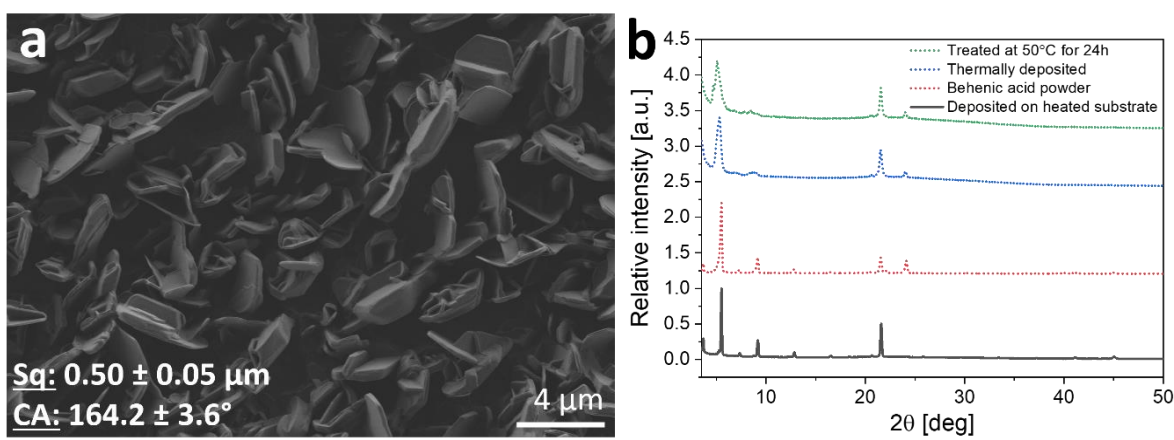

).

SI 2: Characterization of deposited behenic acid on heated glass substrate: a: HR-SEM micrograph of obtained surface, roughness and water contact angle measurements. b: XRD pattern of deposited coating on heated glass substrate (solid line) compared to previously showed powder, coating and heated coating diffractions (dashed lines).

Interestingly, when the deposition was performed upon heating, completely different morphology was obtained: the grown crystals are similar in shape to the crystals of fatty acids from group A, but bigger in size (Figure 1 a, b, c, SI 3 a). Moreover, the crystals also differ in their shape from the crystals of behenic acid coating that was heated at 50°C for 24 h after the deposition (Figure 1d - inset). This experiment shows that not only crystals size can be controlled if the deposition performed on heated substrate, but also their overall shape. Comparing to behenic acid deposited on non-heated substrate, the contact angle was not changed significantly (164.2±3.6° vs. 168.4±4.4°), but the roughness was changed from 1.25μm to 0.5μm on heated substrate, a value which is typical to the roughness values of shorter fatty acid coatings (0.30-0.34μm) (Figure 2c). The XRD pattern also differs from the XRD of behenic acid deposited on unheated substrate as well as from XRD of thermally treated coating: the amorphous hump disappeared as well as the peak at ~24.1° that exist in powder and coatings diffractions (SI 4 b). This support the explanation that the crystals shape changed due to improved crystallinity of the deposited material; an addition of thermal energy can compensate the missing energy for more effective crystallization of longer molecules with higher diffusional barrier for crystallization. However, it is important to note that the heating have to be implemented at the deposition step; when similar temperature was used for thermal treatment (50°C for 24h) the required change was not obtained; the morphological and structural changes of already deposited coating require more energy for molecules reorganization than the energy required at the step of deposition from vapor state.

To this end, different amounts of palmitic acid (16C) and cerotic acid (26C) were deposited on glass substrates for better understanding of film formation process (**Error! Reference source not found.**).

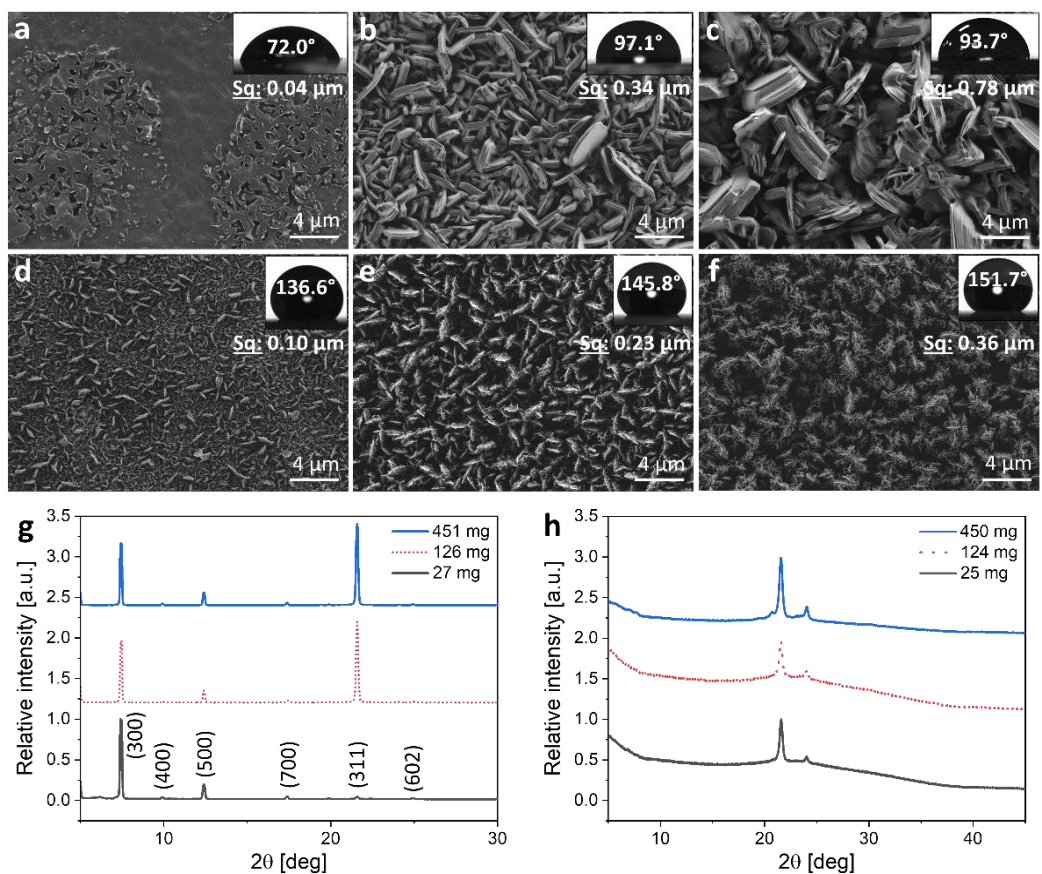

SI 5: Characterization of coatings of different deposited weight of palmitic acid and cerotic acid. a-f) HR-SEM micrographs, insets: contact angles of ethylene glycol and roughness values. a) Palmitic acid, 27 mg, b) Palmitic acid 126 mg, c) Palmitic acid 451 mg, d) Cerotic acid 25 mg, e) Cerotic acid 124 mg, f) Cerotic acid 450 mg. g) XRD of palmitic acid different deposited weight coatings, h) XRD of different deposited weight coatings of cerotic acid.

For palmitic acid (16C), a decrease of deposited weight down to 27 mg resulted with different morphology relatively to thicker coatings: smooth surface with structured areas where a crystals growth is initiated was formed (**Error! Reference source not found.**, a). The same unstructured coating is observed in cross section images of deposited SFAs, mainly in palmitic acid (16C) (Figure 1 a). This initial wetting surface coating exist in all samples and probably provides the base for crystals growth that are formed from afterwards deposited SFAs. As a result of flat non-structured surface, lower ethylene glycol CA (72.0°) and lower roughness (0.04μm) relative to 126 mg deposited coating (CA 131.3° and Sq 0.34μm) were obtained. The difference of the thinner coating can be also seen from the XRD pattern: the intensity of preferred (311) peak of the thicker coatings is significantly lower for the 27 mg deposited coating, indicating the {h00} preferred orientation of the initial wetting layer (SI 5, g). Interestingly, for deposited 451 mg coating of palmitic acid the CA of ethylene glycol decreased relatively to 126 mg coating down to 93.7°, while the roughness increased up to 0.78μm. Based on the HR-SEM micrograph, the developed crystals are coarser; that caused to an increase of the solid-liquid interface and as a result the CA decreased (SI 5, c).

Oppositely to palmitic acid (16C), even a deposition of 5-fold smaller amount (25 mg) of cerotic acid (26C) resulted with characteristic surface morphology, similarly to the observed crystals when higher amount was deposited (SI 5 d, e, f). For this case, the more cerotic acid (26C) was deposited, the more prominent hierarchical structure was developed and correspondingly, higher ethylene glycol contact angle and roughness values were achieved (SI 5 insets in d, e, f). The XRD patterns of cerotic acid coatings remained similar for different deposited weights; the two main peaks exist for all the samples and a minor decrease of the amorphous hump of 450 mg deposited sample indicate higher crystallinity of that coating (SI 5, h).

Overall, deposition of longer SFAs can provide better superhydrophobicity and more prominent hierarchical structure when smaller material amount is used relative to shorter SFAs.

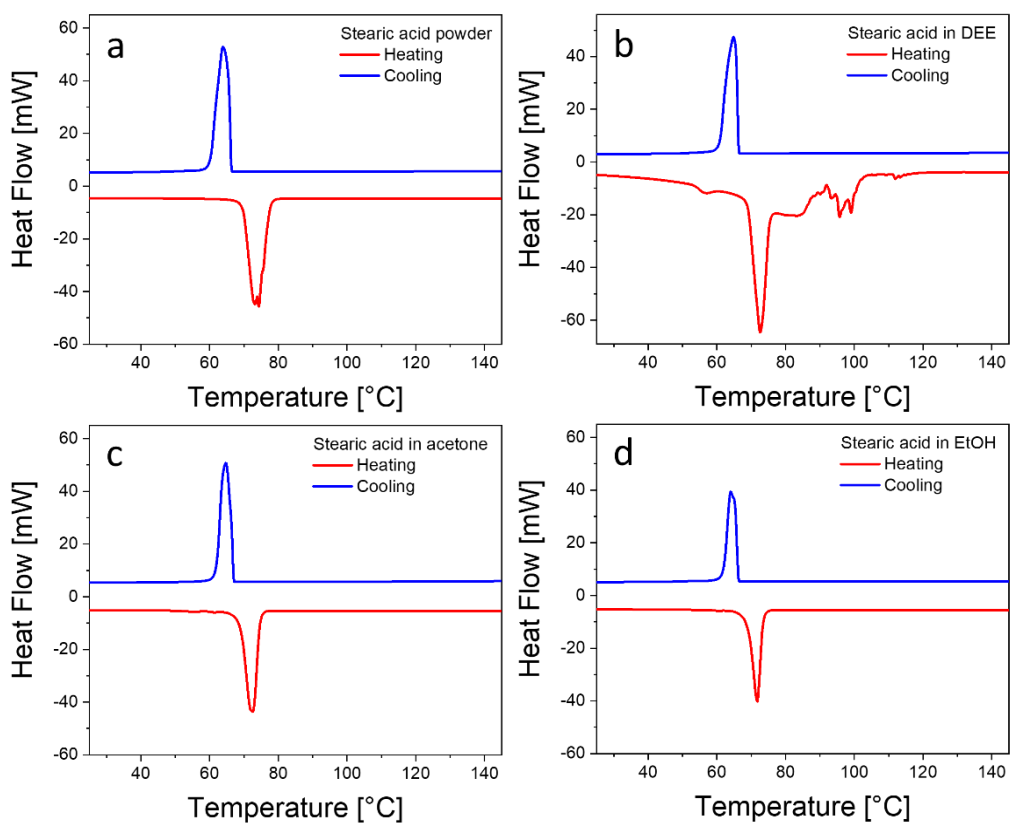

SI 6: DSC measurements of non-processed powder and of different solvent-based sprayed stearic acid (18C). a. non-processed powder, b. diethyl-ether-sprayed, c. acetone-sprayed, c. ethanol-sprayed.

DSC of the coating materials was performed. The DSC of unprocessed stearic acid, acetone- and ethanol-based sprayed stearic acid show curves that comprise single endothermal and exothermal peaks at the heating and cooling curves respectively, proving the conclusion from heat-treated samples XRD that the left-shifted unexpected peaks origin is in the lattice distortion but not in impurities (SI 6, a, c, d). The heating curve of diethyl ether-based spray contains additional endothermal peaks that indicate a contamination of the sample that could also contribute to the XRD pattern (SI 6, b, Figure 4, a3-c3). However, since similar unexpected peaks appeared at both acetone- and even more significantly in diethyl ether-based sprays, and thermal treatment caused to decrease of these peaks' intensity while the main (300) peak intensity increased, it could be concluded that the main contribution to these peaks appearance is of the lattice distortion.

Characterization of palmitic acid and arachidic acid diethyl ether based sprayed coatings:

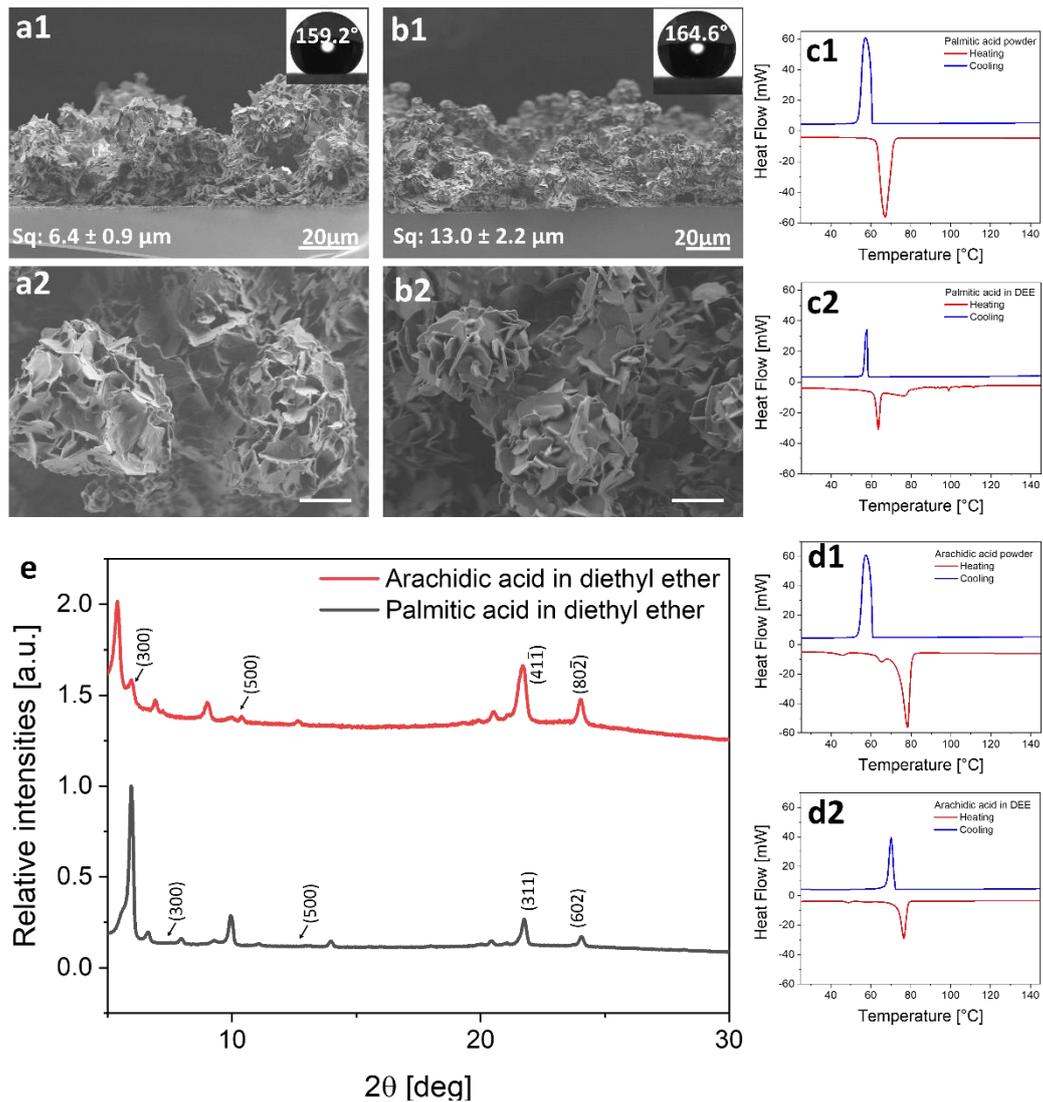

SI 7: Characterization of a. palmitic acid and b. arachidic acid diethyl ether-based coatings: a1-b1 – HR-SEM cross-sectional views. Insets: water CA and roughness values. a2 – b2 – planar views of spray coatings, scale bar is 4µm, c-d - DSC measurements of non-processed powders and diethyl ether-based sprayed SFAs: c1 – palmitic acid powder, c2 – sprayed palmitic acid, d1 – arachidic acid powder, d2 – sprayed arachidic acid. e – XRD of deposited coatings.

SI 7 a – b shows the morphology of palmitic and arachidic acids diethyl ether-based sprayed surfaces, which are very similar to the morphology of stearic acid diethyl ether-based sprayed sample (Figure 4 a). Their surface is covered with aggregates of crystals, which contribute to creating a high surface roughness several microns, respectively, as well as CAs higher than 159° and CAH lower than 6° for both coatings (SI 7 a1, b1). Similar phenomena of unexpected XRD left-shifted peak that was observed for deposited stearic acid (Figure 4 c1) can be seen in SI 7 e. DSC results of unprocessed palmitic acid (SI 7 c1, c2, d1, d2) show endothermal and exothermal single peaks during the heating and the cooling processes, respectively, with few endothermal peaks during the heating of processed sample, similarly to DSC of diethyl ether-processed stearic acid (SI 6 b). Two additional endothermal peaks at lower temperature than the main peak of arachidic acid powder appear (SI 7 d1) and may indicate residual impurities in that powder with no relation to spray processing. These peaks did not appear at cooling process. Cooling DSC curves of all SFAs remained clear, as the cooling curve of stearic acid.

To study the contribution of surface chemistry to the antimicrobial properties of the spray-coated surfaces we eliminated the surface morphology factor by using SFA powders. SFA powders were added to both bacteria culture, incubated for ~24h at 37°C and plated on NB agar for ~24h at 37°C. The results are summarized in the table in SI 8:

SI 8: E. coli and L. innocua count after the incubation with SFA powders in liquid media for 24h at 37°C estimated by the drop-plate method.

| Sample | E. coli [CFU mL$^{-1}$*] | L. innocua [CFU mL$^{-1}$] |
| --- | --- | --- |
| **Control** | $1.2 \pm 0.2 \cdot 10^8$ | $2.4 \pm 0.2 \cdot 10^5$ |
| **Palmitic acid (16C)** | $1.4 \pm 0.6 \cdot 10^8$ | $1.7 \pm 0.5 \cdot 10^4$ |
| **Stearic acid (18C)** | $1.8 \pm 0.05 \cdot 10^8$ | $1.9 \pm 0.5 \cdot 10^4$ |
| **Arachidic acid (20C)** | $9.6 \pm 3.2 \cdot 10^7$ | 0 |

*CFU – Colony-forming unit

Table SI 8 shows the effect of added powders of SFAs on bacterial growth. The results show that the tested SFAs have no inhibiting effect on Gram-negative *E. coli* growth; same order of magnitude of colonies number was obtained when control sample and samples that were incubated with SFAs presence were cultured. Following this result, it can be concluded that the surface morphology has the main contribution to the antimicrobial effect of the sprayed SFAs coatings. However, the same SFAs do have growth inhibition effect on Gram-positive *L. innocua*: one order of magnitude growth inhibition was seen after addition of palmitic acid (16C) and stearic acid (18C) and full growth inhibition was achieved when arachidic acid (20C) powder was added to the samples. This bacterial growth inhibition effect probably contribute to increased deal\live cells ratio that was seen on CLSM images and to the cells morphology change observed on HR-SEM (**Error! Reference source not found.**). However, it is important to notice that when the fatty acid is crystallized into micron-size crystals during the deposition process, the chemical effect may be decreased due to lower availability of the SFA relatively to powdered SFA.